\documentclass{aipproc}
\layoutstyle{6x9}
\def\msun{\mbox{${{\rm M}_\odot}$}}

\def\kms{\mbox{$\;{\rm km\ s}^{-1}$}}     %km s -1%
\def\kmsmpc{\mbox{$\;{\rm km\ s}^{-1}\ {\rm Mpc}^{-1}$}}     %km s -1%

\def\cc{\mbox{$\;{\rm cm}^{-3}$}}
\def\mug{\mbox{$\;\mu$G}}
\def\lesssim{\lower.5ex\hbox{$\; \buildrel < \over\sim \;$}}

\def\ergs{\mbox{$\;{\rm ergs}$}}

\def\etal{{\it {et al.}}}
\def\aap{{\it {Astron. \& Astrophys.}}}

\def\apj{{\it {ApJ}}}

\def\nat{{\it {Nature}}}

\begin{document}
\title{Nonthermal Particles and Radiation Produced by Cluster Merger Shocks}

\author{Robert C.\ Berrington}{
address={ASEE Postdoctoral Fellow, 
Naval Research Laboratory, Code 7653, Washington, DC 20375-5352},
email={rberring@gamma.nrl.navy.mil}
}

\author{Charles D.\ Dermer}{ address={Naval Research Laboratory, Code 7653,
Washington, DC 20375-5352},email={dermer@gamma.nrl.navy.mil} }

\begin{abstract}
We have developed a numerical model for the temporal evolution of particle and
photon spectra resulting from nonthermal processes at the shock fronts formed
in merging clusters of galaxies. Fermi acceleration is approximated by
injecting power-law distributions of particles during a merger event, subject
to constraints on maximum particle energies. We consider synchrotron,
bremsstrahlung, Compton, and Coulomb processes for the electrons, nuclear,
photomeson, and Coulomb processes for the protons, and knock-on electron
production.  Broadband radio through $\gamma$-ray light curves radiated by
nonthermal protons and primary and secondary electrons are calculated both
during and after the merger event. Using ROSAT observations to establish
typical parameters for the matter density profile of clusters of galaxies, we
find that merger shocks are weak and accelerate particles with relatively soft
spectra.  Our results suggest that only a minor contribution to the diffuse
extragalactic $\gamma$-ray background can originate from cluster merger
shocks.
\end{abstract}

\maketitle

\section{Introduction}

In the hierarchical merging cluster scenario, clusters grow in mass by
accreting nearby clusters.  Approximately 30--40\% of galaxy clusters show
evidence of substructure in both the optical \cite{beers:82} and X-ray
wavelengths \cite{forman:81}.  Velocity differences between the observed
structures is $\approx\!\!1000$--$2000$\kms.  With gravitational forces
driving the interaction between the two systems, cluster mergers are
consistent with highly-parabolic orbits.  Typical sound speeds within the
intracluster medium (ICM) are $\approx\!\!1000$\kms, so shocks will form at
the interaction boundary of the two systems.

Shock fronts that form in the ICM as a result of a cluster merger event are
thought to be associated with the cluster {\em radio relics}, which are
diffuse emission found on the cluster periphery with no known optical
counterpart.  The shock compression will orient any existing cluster magnetic
field into the plane of the shock.  Radio relics are characterized by highly
organized magnetic fields with field strengths in the $\sim\!\!1$\mug\ range
with linearly polarized field lines in the vicinity of the shock
\cite{ensslin:98}.  The shock front will accelerate a fraction of the thermal
particles within the ICM by first-order Fermi acceleration.

It has been proposed that cluster mergers are the dominant contributor to the
diffuse $\gamma$-ray background \cite{loeb:00}.  Some unidentified EGRET
sources are claimed to be associated with $\gamma$-ray emission from galaxy
clusters \cite{totani:00}.  Excess EUV emission from Coma can be explained by
nonthermal electrons accelerated at merger shocks \cite{atoyan:00}.
Variations in radio surface brightness will result from superposition of
cluster emissions \cite{waxman:00}.

\section{Models}

We present the results of a computer code designed to calculate the
time-dependent particle distribution functions evolving through adiabatic and
radiative losses for electrons and protons accelerated by the first-order
Fermi process at the cluster merger shock \cite{berrington:02}.  The model
calculates the shock speed from the shock formed in a cluster merger event by
the interaction speeds expected from two point masses interacting under their
mutual gravity.  The point masses are assumed to be on elliptical orbits whose
onset is initiated by the collapse and merger of primordial density
fluctuations.  The collision is a result of the two bodies ``falling out'' of
the Hubble flow and onto a nearby cluster.  Expected total orbital energies
are $E_{\rm tot}\sim10^{63}$--$10^{64}\ergs$.  Expected collision velocities
between a $10^{15}\msun$ and a $10^{14}\msun$ mass cluster range from
$\sim\!\!\!1800$--$3500\kms$.  The electron and proton distribution functions
originate from a momentum power-law injection spectrum.  In terms of particle
kinetic energy $E_{e,p} = (\gamma_{e,p} - 1) m_{e,p} c^{2}$, the injection
function is
\begin{equation}
Q_{e,p}(E_{e,p},t) = Q^{0}_{e,p} c^{s(t) - 1} \left[ E_{e,p} \left( E_{e,p} +
  2 m_{e,p} c^{2} \right) \right]^{-\frac{s(t)+1}{2}} \left( E_{e,p} + m_{e,p}
  c^{2} \right) e^{-\frac{E_{e,p}}{E_{\rm max}(t)}}\; ,
\label{eqn:power-law}
\end{equation}
where an exponential cutoff $E_{\rm max}$ has been applied and is determined
by the maximum energy associated with the available time since the beginning
of the merger event, by a comparison of the Larmor radius with the size scale
of the system, and by a comparison of the energy-gain rate through first-order
Fermi acceleration with the energy-loss rate due to adiabatic, synchrotron and
Compton processes. The injected particle function is normalized through the
normalization constant $Q_{e,p}^{0}$ which is determined by
\begin{equation}
\int_{E_{\rm min}}^{E_{\rm max}}\ dE\ E_{e,p} Q_{e,p}(E,t) =
\frac{\eta_{e,p}}{2} A_{s} \eta^{e}_{\rm He} \langle n \rangle m_{p}
v_{s}^{3}\;.
\label{eqn:normalization}
\end{equation}
where $\eta_{e,p}$ is an efficiency factor, $v_{s}$ is the shock speed $A_{s}$
is the area of the shock front; $E_{\rm max}$ and $E_{\rm min}$ are the
maximum and minimum particle energies; $\eta^{e}_{\rm He}$ is an enhancement
factor to account for the presence of ions heavier than Hydrogen.  We assume
an efficiency factor $\eta_{e,p} = 5\%$ for both protons and electrons.  With
this method the total energy deposited into nonthermal particle production is
$\eta_{e,p}E_{\rm tot}$.  The particle density $\langle n\rangle$ is
calculated according the $\beta$ model, and averaged over the shock front.

The time evolving particle spectrum is determined by solving the Fokker-Planck
equation in energy space for a spatially homogeneous ICM, given by
\begin{equation}
\frac{\partial N(E,t)}{\partial t} = \frac{1}{2} \frac{\partial^{2}}{\partial
E^{2}}[D(E,t) N(E,t)] - \frac{\partial}{\partial E}[\dot{E}_{\rm tot}(E,t)
N(E,t)] + Q(E,t) - \sum_{i=\pi,p\gamma,d} \frac{N(E,t)}{\tau_{i}(E,t)}\;.
\label{eqn:Fokker-Planck}
\end{equation}
The quantity $\dot{E}_{\rm tot}(E,t)$ represents the total synchrotron,
Compton, Coulomb, and adiabatic energy-loss rate for electrons, and the sum of
the Coulomb and adiabatic energy-loss rates for protons.  Both protons and
electrons are subject to diffusion in energy space by Coulomb interactions.
The protons experience catastrophic losses due to proton-proton collisions on
the timescale $\tau_{\pi}$, proton-$\gamma$ collisions on the timescale
$\tau_{p\gamma}$, and spatial diffusion from the cluster on timescale
$\tau_{d}$.  Spectra of secondary electrons and positrons are calculated from
pion-decay products, and are subject to the same physical processes as the
primary electrons.  $Q(E,t)$ is the particle injection function.

The synchrotron, Compton, bremsstrahlung, and pion-decay $\gamma$-ray spectral
components are calculated from the particle spectra following the methods
described by \cite{berrington:02}.  We use a standard parameter set with a
mean ICM number density $n_{0} = 10^{-3}\cc$ and a uniform cluster magnetic
field of $B = 1.0\mug$. Thus, $V(t) = A(t) v_s t$.  

\begin{figure}[t]
%\spaceforfigure{5in}{5in}
\includegraphics[width=6.0in]{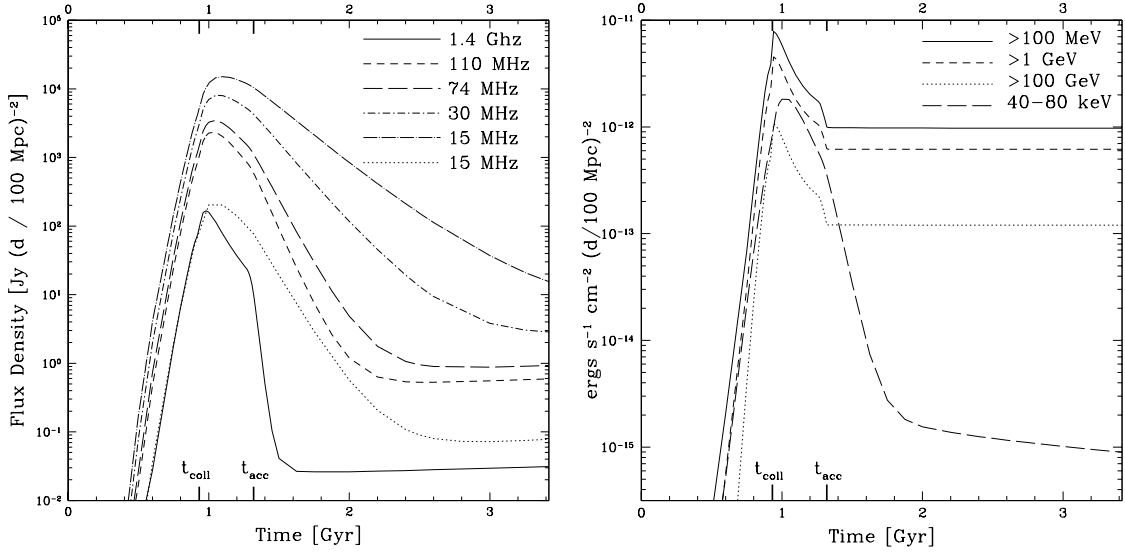}
\caption{Light curves at various observing frequencies produced by a shock
formed in a merger between $10^{14}\msun$ and $10^{15}\msun$ clusters that
begins at $z_{i}$=0.3 and is evolved to the present epoch (t=3.42 Gyr). All
light curves are for a magnetic field strength of B=$1.0\mug$ except the 15
MHz light curve is also calculated with a magnetic field strength of
B=$0.1\mug$ (dotted curve).  Radio light curves in Jansky units are given at
15 MHz, 30 MHz, 74 MHz, 110 MHz, and 1.4 GHz on the left panel, and light
curves in energy flux units are given at 40-80 keV, $>$100 MeV, $>$1 GeV and
$>$100 GeV in the right panel.  }
\label{fig:light_curves}
\end{figure}

\section{Results}

Fig.\ \ref{fig:light_curves} shows the light curves at various energies for a
cluster merger shock between a $10^{14}\msun$ and $10^{15}\msun$ with radii of
1.5 Mpc and 0.75 Mpc, respectively, with a central gas number density of
$n_{0}$=$10^{-3}$, and a magnetic field of B=1\mug\ is assumed.  The assumed
core radius for the dominant cluster profile is 250 kpc, and $\beta$=0.75.  We
assume a cosmology defined by ${\rm H}_{0}$ = 70\kmsmpc, and
($\Omega_{0}$,$\Omega_{R}$,$\Omega_{\Lambda}$)=(0.3,0.0,0.7).  We assume the
shock forms at a redshift of $z_{i}$=0.3.  The light curves exhibit a similar
characteristic independent of frequency with the peak luminosity occurring
when the centers of mass of the two clusters pass.  The emission slowly decay
and approach a plateau at times $t>t_{\rm acc}$ when particle injection has
stopped.  The rate of decay slows with decreasing energy with the lowest
energies exhibiting the slowest decay rates.  The late time plateaus are from
$\pi^{0}$ $\gamma$-rays for photons with E$>$1 GeV, and secondary electrons
for the radio energies.  With these expected fluxes, it is unlikely that the
more than a few of the isotropic unidentified EGRET sources can be attributed
to the radiation of nonthermal particles produced in cluster merger shocks.

\begin{figure}[t]
%\spaceforfigure{5in}{5in}
\includegraphics[width=3.15in]{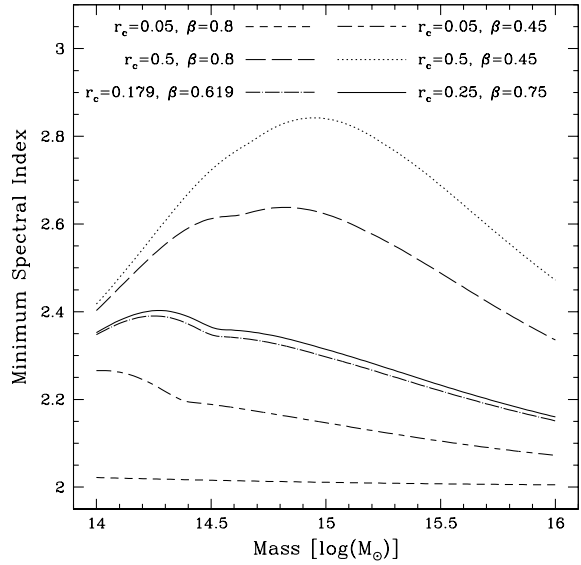}
\caption{Calculations of the hardest particle injection spectral indices
$s_{\rm min}$ formed in cluster merger shocks as a function of the dominant
cluster mass, for various values of $r_{c}$ and $\beta$.  The mass of the
merging cluster is assumed to be $10^{14}\msun$.  Our adopted values are
plotted by the solid line, and dot-dashed line shows $s_{\rm min}$ plotted
against the dominant cluster mass for the average values of $r_{c}$ and
$\beta$ obtained by ROSAT observations of 45 Abell clusters \cite{wu:00}.}
\label{fig:minimum_spectral_index}
\end{figure}

The diffuse extragalactic $\gamma$-ray background is a featureless power law
with photon index of 2.10($\pm$0.03).  Fig.\ \ref{fig:minimum_spectral_index}
shows the hardest particle injection spectral index, $s_{\rm min}$, for a
cluster merger shock as function of the mass of the dominant cluster with
varying values of core radii, $r_{c}$, and $\beta$'s.  The particle spectral
indices expected from the average values of $r_{c}$ and $\beta$ obtained from
ROSAT observations of 45 Abell clusters \cite{wu:00} are 2.2-2.4.  Only the
dark matter profiles with strong central peaks produce particle indices
$<$2.1.  This discrepancy in the particle indices suggests that nonthermal
radiation from cluster merger shocks can make only a minor contribution to the
diffuse extragalactic $\gamma$-ray background unless dark matter halos contain
strong central peaks.

\end{document}